\documentclass[fleqn,10pt]{wlscirep}
\usepackage[utf8]{inputenc}
\usepackage[T1]{fontenc}
\usepackage{setspace}
\usepackage{upgreek}

\newcommand{\ket}[1]{{\left| {#1} \right\rangle}}

\renewcommand{\o}{\textit{o}}
\newcommand{\m}{\textit{m}}
\newcommand{\g}{\textit{g}}
\newcommand{\omg}{\textit{omg}}

\title{\omg\ Blueprint for trapped ion quantum computing with metastable states}

\author[1]{D. T. C. Allcock}
\affil[1]{Department of Physics, University of Oregon, Eugene, OR, USA}
\author[2,3,4]{W. C. Campbell}
\affil[2]{Department of Physics and Astronomy, University of California Los Angeles, Los Angeles, CA, USA}
\affil[3]{Center for Quantum Science and Engineering, University of California Los Angeles, Los Angeles, CA, USA}
\affil[4]{Challenge Institute for Quantum Computation, University of California Los Angeles, Los Angeles, CA, USA}
\author[5,6]{J. Chiaverini}
\affil[5]{Lincoln Laboratory, Massachusetts Institute of Technology, Lexington, MA, USA}
\affil[6]{Massachusetts Institute of Technology, Cambridge, MA, USA}
\author[7]{I. L. Chuang}
\affil[7]{Center for Ultracold Atoms, Department of Electrical Engineering and Computer Science, Department of Physics,
Massachusetts Institute of Technology, Cambridge, MA, USA}
\author[2,3,4]{E. R. Hudson}
\author[1]{I. D. Moore}
\author[2, 8]{A. Ransford}
\affil[8]{Honeywell Quantum Solutions, Broomfield, CO, USA}
\author[2, 8]{C. Roman}
\author[5,6]{J. M. Sage}
\author[1]{D. J. Wineland}

\begin{abstract}
Quantum computers, much like their classical counterparts, will likely benefit from flexible qubit encodings that can be matched to different tasks.  
For trapped ion quantum processors, a common way to access multiple encodings is to use multiple, co-trapped atomic species.
Here, we outline an alternative approach that allows flexible encoding capabilities in single-species systems through the use of long-lived metastable states as an effective, programmable second species.  
We describe the set of additional trapped ion primitives needed to enable this protocol and show that they are compatible with large-scale systems that are already in operation.

\end{abstract}
\begin{document}

\flushbottom
\maketitle

\thispagestyle{empty}

\section{Introduction}

Trapped atomic ions manipulated with electromagnetic fields show great promise for large scale quantum computation~\cite{BruzewiczMITLLIonReview2019}, yet existing systems are error prone and not of a scale compatible with known practical applications.  Capabilities required to realize this promise likely include accomplishing the following, in arrays of many ions and while maintaining the high fidelities obtained in few-ion experiments:  low-crosstalk addressing of individual qubits; in-algorithm readout of ancillas with minimal decoherence of unmeasured ion-qubits; real-time production of photon-mediated entanglement between sub-arrays; and fast, low-motional-state-excitation transport and reordering of at least a subset of ions within a quantum register to enable high connectivity and to isolate motional mode frequencies of multi-ion crystals.

While these requirements form a daunting list, an architectural methodology that has the potential to address all of these challenges has begun to take shape, and many of the related proof-of-principle experiments have been performed---this is the dual-species scheme~\cite{Wineland1998ExperimentalIssues}.  This versatile approach employs an additional set of ions of a second atomic species to perform likely necessary functions, such as sympathetic cooling and mapped ancilla readout, that will result in deleterious scattered light between ions of the same species.  In this case, the large energy differences between transitions in the two ion species effectively isolate the qubits stored in electronic states of one species from the light scattered by the other.

While these dual-species systems have been utilized in many basic demonstrations of the functionality described above, they bring significant challenges of their own, such that new roadblocks emerge during their use, both fundamental and technical.  The former include the difficulties associated with:  cooling all shared motional modes due to decoupling by species~\cite{Wubbena2012Sympathetic,Sosnova2020Character}; differential sensitivity to stray static fields, gradients in the pondermotive-trap potential, and trap anharmonicities \cite{Barrett2003Sympathetic,Home2011Normal}; and the transport of multi-species ion crystals without significant motional excitation~\cite{Palmero2014Shuttle}.  All of these are due to the mass
mismatch of (singly ionized) atoms of different elements (see Ref.~\cite{HomeMixedSpecies2013} for a review).  The technical challenges include the requirement of almost twice as many stabilized laser systems and optics as single-species systems, the potential need for complex interspecies operations~\cite{TanBeMg2015,BruzewiczSrCa2019,HughesCaSr2020}, and for ground-state hyperfine qubits (with the exception of Ba$^+$), the requirement to have significant laser power in the blue or UV for high-fidelity quantum logic.  Dual-species systems thus often trade one set of difficult challenges for another.

Here we propose an alternate architecture for solving the problems addressed by dual-species systems, taking the best aspects of the two-species methodology while eliminating, by construction, many of the issues that emerge with the introduction of a second atomic species (see also \cite{yang2021realizing}).  In our approach, the simultaneous use of multiple types of electronic qubits within a single species of ion, along with interconversion between the qubit types, can allow for two spectrally separate and dynamically configurable registers of qubits in a collection of trapped singly-ionized atoms of the same species and hence the same mass.
Such an architecture has dual-species functionality in a single-species array, a potentially  powerful combination.

\section{The \omg\ architecture concept}

\begin{figure}[t]
\begin{centering}
\includegraphics[width=0.75\columnwidth]{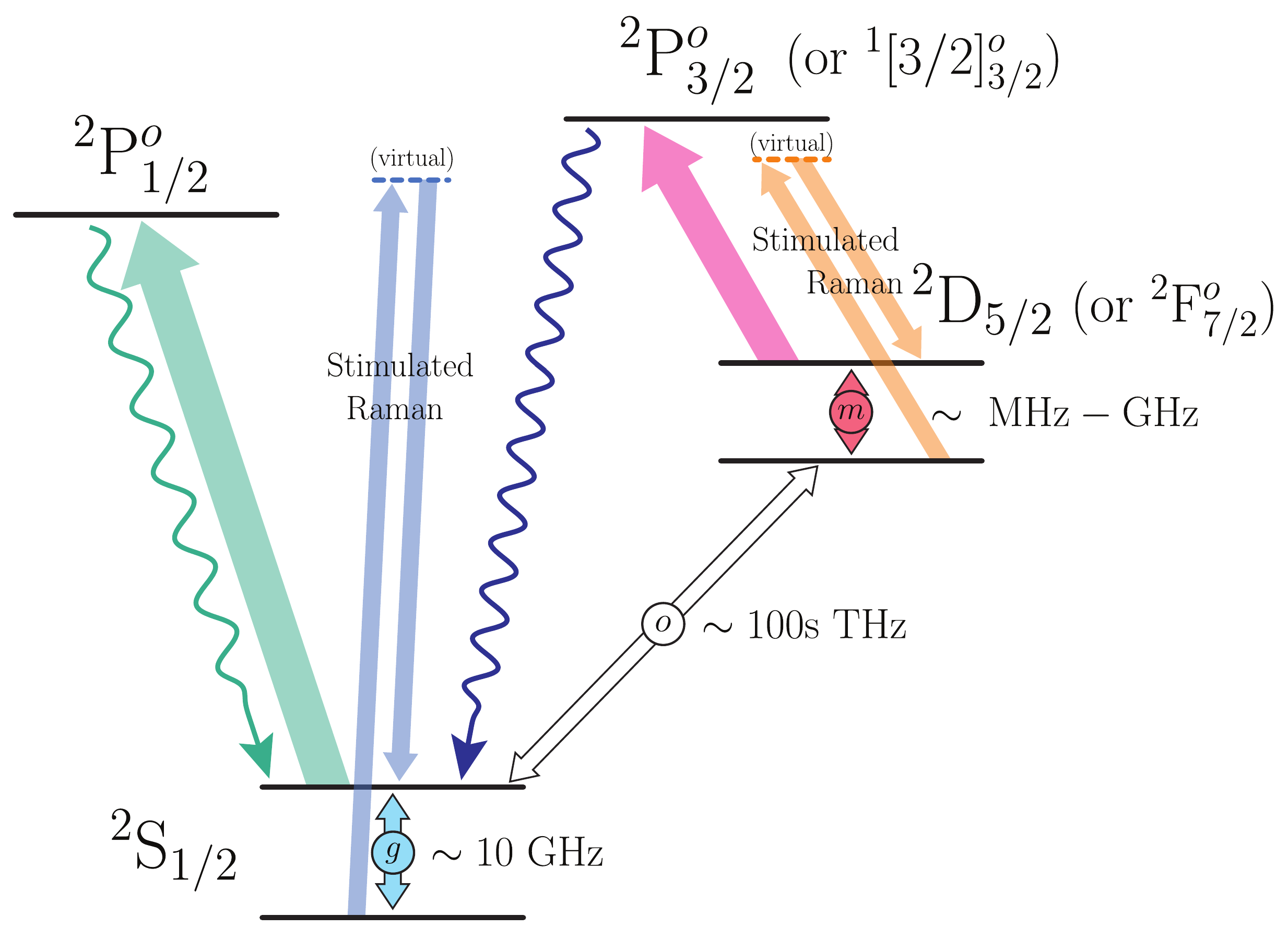}
\caption{\label{fig:LevelDiagram}Common structure of alkaline earth ions with hyperfine structure and metastable electronic states showing encoding of three types of qubits with relevant transitions for the tasks described in Fig.~\ref{fig:OMGModes}.  The fast cycling transition on ${}^2\mathrm{P}_{1/2}^o \leftrightarrow {}^2\mathrm{S}_{1/2}$ is used for open channel processes (laser cooling, state preparation, and readout), including heralded state preparation of the \m\ qubit by optical pumping of one qubit state via the highest shown excited state (\S \ref{sec:StatePrep}).
The \o\ qubit transition is not electric dipole (E1) allowed, and typically requires a narrow linewidth laser (\S \ref{sec:CoherentConversion}).
} 
\end{centering}
\end{figure}

We term this new architecture as \omg, after the three types of electronic qubits employed: optical-frequency (\o), metastable-state (\m), and ground-state (\g) qubits. These qubit types are shown schematically in Fig.~\ref{fig:LevelDiagram}, and can be housed in a single ion species.  
The \o\  qubit is composed of one ground state and one metastable state, with an energy splitting corresponding to an optical frequency.
The \m\  qubit is composed of two metastable atomic states, such as hyperfine or Zeeman levels, of an atomic ${}^2\mathrm{D}_{5/2}$ or ${}^2\mathrm{F}^o_{7/2}$ state.
In order to be useful for a qubit, such metastable states must have lifetimes that are long compared to the time quantum information is processed or stored in them, but they need not be as long as the lifetimes of \g\  qubit states. 
The \g\ qubit is composed of two very long-lived ground states (once again, hyperfine or Zeeman levels) of an atomic ${}^2\mathrm{S}_{1/2}$ state manifold. 
Interconversion between qubit types (type casting) is accomplished by driving transitions between ground and excited electronic states.  

From a practical viewpoint, the \o, \m, and \g\  qubits have a variety of lifetimes and wavelengths (see Table \ref{table:IonSpecies}), and this leads to trade offs in utilizing the architecture for quantum computation. Some aspects of this architectural concept have been demonstrated previously; coherent conversion between \o\ and \m\ qubits is well established~\cite{Roos2004Shelve}, single-qubit \m\ gates have been performed~\cite{Sherman2013}, and very recently the two capabilities have been combined~\cite{yang2021realizing}.  However \m\ qubits are less developed than \o\ and \g\ qubits, with two-qubit \m\ gates in particular yet to be demonstrated.  In the remainder of this section, we first discuss the general benefits of the \omg\ architecture, then summarize the options available for ground and metastable states in particular species of ions, and detail three potential modes for using \omg\ qubits architecturally.

\subsection{Benefits of the \omg\ architecture}
In this architecture, with judicious choice of ground and metastable states for the \o, \m, and \g\  qubits, all of the required capabilities for trapped-ion quantum computing described above can be implemented with just a single species of ion in the quantum register.  Furthermore, like in the dual-species architecture, these tasks can be performed by a subset of the register with vanishing impact on other ions.  This crucially allows for state preparation (which includes sympathetic laser cooling), measurement, calibration, and photon-assisted entanglement, all during execution of a quantum algorithm. The decoherence induced by scattered light renders such mid-algorithm operation impractical in typical single-ion-species systems where ions are spaced closely together.

Though a dual-species approach can provide this capability, the benefits of the \omg\ architecture over the dual-species architecture are many.  In particular, a register of identical ions offers several advantages:  laser cooling is more efficient; motional modes couple to differently encoded ions indiscriminately; the particular spatial ordering of differently encoded ions in an ion crystal does not affect the motional mode structure, thus making the system significantly less sensitive to ion re-ordering events; the deleterious impact from stray electric fields and pseudopotential gradients is reduced; transport of multi-ion crystals is greatly simplified; and mid-algorithm calibration of the system using dedicated calibration ions, fundamentally identical to the data ions, can potentially reduce systematic error when compared with calibrating with ions of a different species.  The \omg\ architecture also allows for dynamic reconfiguration of the positions of different types of qubits in an ion crystal via fast, site addressed laser-based interconversion operations as opposed to relatively slower shuttling-based approaches.  In addition the required lasers and optical components are reduced when compared to the dual-species approach, and only a single species needs to be loaded.

\subsection{Physical states available for the \omg\ architecture}

Most of the alkaline-earth-like species used for trapped ion quantum computation have easily accessible metastable $^2\mathrm{D}_{5/2}$ or ${}^2\mathrm{F}_{7/2}^o$ states that are not populated by the cycling transitions from the ground-state used for cooling and readout, with lifetimes ranging from hundreds of milliseconds to years, as summarized in Table~\ref{table:IonSpecies}.  All of these species have isotopes with nonzero nuclear spin, which leads to hyperfine splitting and access to `clock' qubits, which have zero first-order magnetic field sensitivity in both the ground and metastable manifolds.

\begin{table}[htbp]
    \centering
    \begin{tabular}{c|c|c|c|c|c|c|c|c}
    \hline \hline
        Species & $I$ & \g\ qubit & \g\ qubit & \m & \m & \m\ qubit & \m\ qubit & \o\ qubit \\
         &  & $F\leftrightarrow F'$ & splitting & state & lifetime &  $F\leftrightarrow F'$ & splittings (MHz) & wavelength\\
        \hline
         ${}^{43}\text{Ca}^+$ & 7/2 & 3$\leftrightarrow$4 & 3.2\,GHz & $^2\mathrm{D}_{5/2}$ & 1.2\,s & 1$\leftrightarrow$2,..., 5$\leftrightarrow$6 & 7, 10, 15, 20, 25 & 729\,nm \\
         ${}^{87}\text{Sr}^+$ & 9/2 & 4$\leftrightarrow$5 & 5.0\,GHz & $^2\mathrm{D}_{5/2}$ & 0.39\,s & 2$\leftrightarrow$3,..., 6$\leftrightarrow$7 & 8.2, 5.2, 2.7, 17, 38 & 674\,nm\\
         ${}^{133}\text{Ba}^+$ & 1/2 & 0$\leftrightarrow$1 & 9.9\,GHz & $^2\mathrm{D}_{5/2}$ & 30\,s & 2$\leftrightarrow$3 & 89 & 1.76\,$\upmu$m \\
         ${}^{135}\text{Ba}^+$ & 3/2 & 1$\leftrightarrow$2 & 7.2\,GHz & $^2\mathrm{D}_{5/2}$ & 30\,s & 1$\leftrightarrow$2,..., 3$\leftrightarrow$4 & 52, 50, 12 & 1.76\,$\upmu$m \\
         ${}^{137}\text{Ba}^+$ & 3/2 & 1$\leftrightarrow$2 & 8.0\,GHz & $^2\mathrm{D}_{5/2}$ & 30\,s & 1$\leftrightarrow$2,..., 3$\leftrightarrow$4 & 72, 63, 0.49 & 1.76\,$\upmu$m \\
         ${}^{171}\text{Yb}^+$ & 1/2 & 0$\leftrightarrow$1 & 12.6\,GHz & $^2\mathrm{F}^o_{7/2}$ & 1.58 years & 3$\leftrightarrow$4 & 3620  & 467\,nm \\
         ${}^{173}\text{Yb}^+$ & 5/2 & 2$\leftrightarrow$3 & 10.5\,GHz & $^2\mathrm{F}^o_{7/2}$ & days-years & 1$\leftrightarrow$2,..., 5$\leftrightarrow$6 & 260, 1000, 130, 920, 3300  & 467\,nm \\
         \hline \hline
    \end{tabular}
    \caption{Lifetimes, frequencies, and states for \omg\ qubits in seven commonly employed trapped ion species for quantum computation, including optical-frequency qubits between single quantum states in the ground and metastable manifolds (\o), metastable-state hyperfine qubits (\m), and ground-state hyperfine qubits (\g).
    Qubit splittings for \g\ and \m\ are given at zero magnetic field. \cite{Benhelm2008Precision,Barwood2003Observation,ChristensenHF2019,Becker1981Precise,Silverans1986Hyperfine,Dzuba2016HyperfineInduced,Xiao2020Hyperfine,Lange2021Lifetime}
    \label{table:IonSpecies}}
\end{table}

Given this structure, a natural question arises: how do these physical state options map onto high performance quantum information storage, logic gates, and state preparation?  From an architectural perspective, attractive states for qubit storage have long lifetimes; ground states are well known to have the longest coherence lifetimes, but as Table \ref{table:IonSpecies} shows, metastable states can also have usefully long lifetimes, particularly in $\mathrm{Yb}^+$.  
For logic gates, speed can be limited by energy splittings (smaller splittings  take longer to resolve), but gate fidelity can also depend on these splittings, and especially the distribution of energy levels around qubit transitions.  
Good gate operations also depend on having technologically convenient qubit wavelengths.
For example, due to the level structure of the species of interest, \m\  qubits could be competitive with \g\ and \o\  qubits for quantum logic by enabling Raman-based operations, for similar intensities, at longer, more technologically favorable wavelengths. 
For state preparation, much depends on available laser cooling mechanisms; \o\  qubits can be most straightforward to prepare and read out, due to their observable fluorescence.
Similarly, species with $I = 1/2$ also provide fast, high-fidelity state preparation of \g\ qubits due to the presence of a frequency-resolved $F = 0$ qubit state. 
In the next section, we consider three distinct schemes for combining these features that leverage the varying properties of the available host atomic ions. 
\subsection{Three modes for the \omg\ architecture}

Three key quantum computation needs are state preparation, gates, and storage.  The \omg\ architecture allows qubit types to be assigned to optimize each of these needs. For example, \g\  and \m\  qubits are largely unperturbed by the laser light used to manipulate the other.  This is particularly important for the dissipative processes such as laser cooling and state preparation, and is essentially responsible for the high fidelity that can be achieved for state measurement of \o\  qubits.  As a result, physical separation in real space, often achieved via ion shuttling in a multi-zone trap (or, alternatively, interspecies gates) may not be needed as frequently if the sensitivity of individual qubits to the applied light can be controlled sufficiently.  The \o\m\g\ platform, then, allows some resource intensive operations to be replaced by potentially easier manipulations in small crystals.  Operations involving ions in separate crystals will then be done via shuttling, but the ease with which any ion in the register can serve as a mass-matched refrigeration ion is expected to reduce the overhead of shuttling.

\begin{figure}[htbp]
\begin{centering}
\includegraphics[width=1\columnwidth]{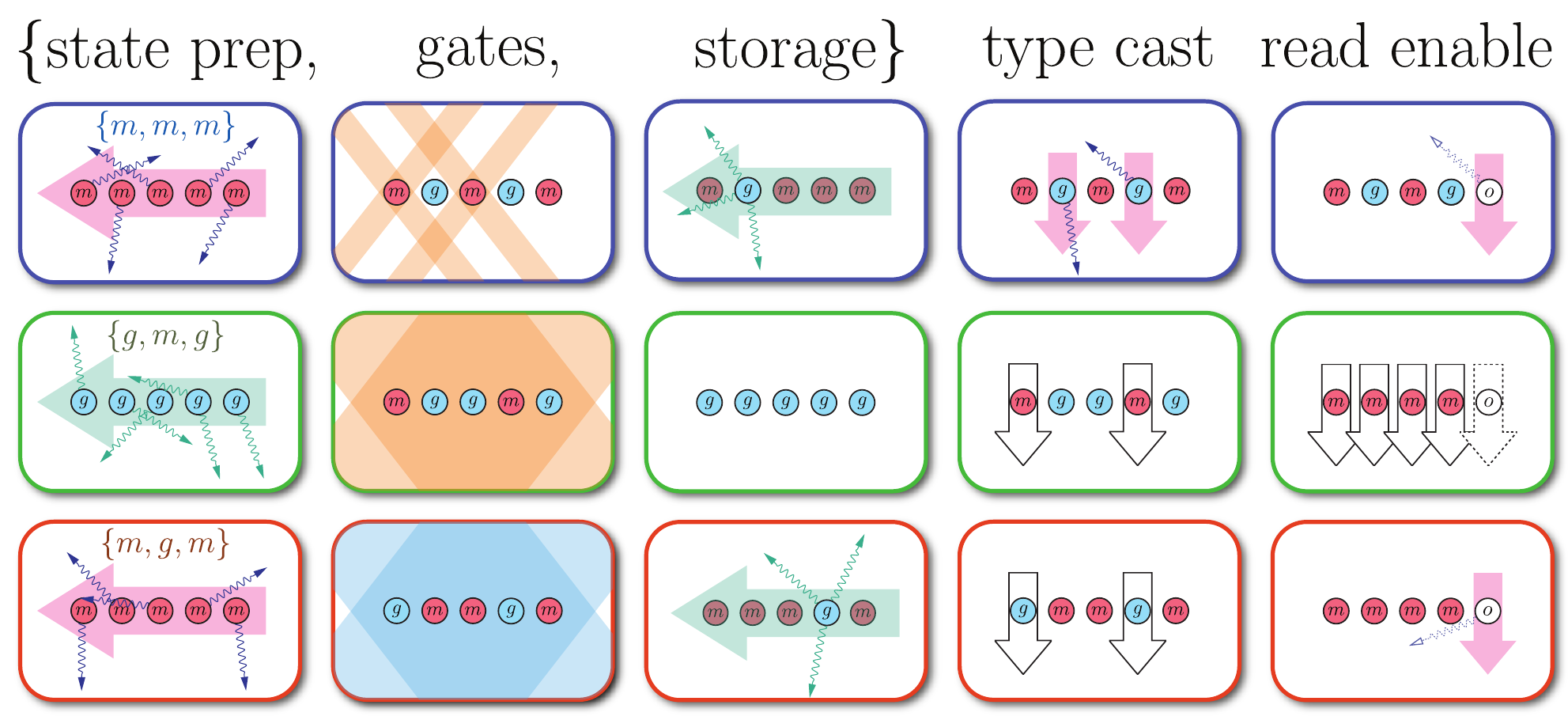}
\caption{\label{fig:OMGModes}Three modes (see \S\ref{sec:mmmMode}-\ref{sec:mgmMode}) for utilizing \o, \m, and \g\  qubits in a small linear ion crystal.  The panels of each row constitute a distinct mode. Modes are designated according to the type of qubit used for \{state preparation (\S\ref{sec:StatePrep}), gates (\S\ref{sec:Gates}), storage\}. Each circle represents an ion, and one qubit of a specific type is embodied by each ion.    In all three cases, cooling is done in \g\ type ions and readout (\S\ref{sec:StatePrep}) in \o\ type manifolds (see also Fig.~\ref{fig:LevelDiagram}).  Block arrows indicate laser beams, and wiggly arrows denote spontaneously scattered photons, with dash arrows denoting conditional transitions.  Laser cooling of \g\ type ions is shown during storage where possible.  Type cast denotes conversion between \o, \m, and \g\ qubits (\S\ref{sec:CoherentConversion}), with open channel conversion accompanied by a spontaneously emitted photon.  Read enable refers to the conversion to \o\ encoding so that subsequent laser interrogation produces state-dependent fluorescence (\S\ref{sec:StatePrep}). Read enable can be done through an open channel in the top and bottom modes, in which case the spontaneously emitted photon is only present if the qubit projects onto the particular state being optically pumped.} 
\end{centering}
\end{figure}

Below, we present three potential architectural modes (depicted in Fig.~\ref{fig:OMGModes}) for utilizing the three qubit types in small crystals, codified by an ordered triple $\{q_1,q_2,q_3\}$ that identifies which type of qubit ($q_i=$ \o, \m, or \g) is to be used for: $\{$state preparation, gates, storage$\}$.  In each case, sympathetic cooling of multi-ion crystals can be done with \g\ type ions while information is protected in the \m\ qubit, and active reconfiguration of the ion types to optimize the cooling can be done without shuttling. Likewise, readout can be performed by encoding the desired information in \o\ qubits, measured via simple manifold discrimination, while information in nearby ions is protected by encoding it in \m\ qubits.  For each mode, we describe the setting, then discuss strengths and weaknesses.  The required new primitive quantum computation operations are then detailed.

\subsubsection{\{\m,\m,\m\} mode} \label{sec:mmmMode}
{\bf Setting: Fig.~\ref{fig:OMGModes} (top).} Since \g\ type ions can be coupled strongly to open channels for cooling, state preparation, and readout via spontaneous photon emission, a natural way to utilize the three qubit types is to confine most of the unitary operations to \m\ qubits.
Here, \m\ qubit gates would be individually addressed by laser beams focused on the desired ions to drive stimulated Raman transitions, similar to \g\ qubits.
Site-specific readout can be effected by individually-addressed read enabling by conversion to \o\ type followed by resonance fluorescence detection.

{\bf Strengths:} One of the strengths of the $\{\m,\m,\m\}$ mode is that no coherent transfer between the different qubit types is required as all coherent operations act on \m\ qubits only.  
This means the technical requirements on any \o\ qubit laser (\textit{e.g.}, a narrow-band laser for directly driving the electric quadrupole (E2) or electric octupole (E3) transition between the ground and metastable states) and the sensitivity to ion motional effects are much reduced.  Indeed, it may even be possible to replace direct \m\ $\rightarrow$ \g/\o\ transitions for type casting entirely with simple and robust optical pumping through electric dipole (E1) transitions to auxiliary excited states.  Also, a unique feature of \{\m,\m,\m\} mode in comparison to those that follow is the fact that laser cooling and \g\ qubit state preparation and readout can be accomplished during single-qubit gates operating on other ions in the same crystal.

{\bf Weaknesses:} The main disadvantages of \{\m,\m,\m\} mode are that storage is limited by the \m\ lifetime (see Table\ \ref{table:IonSpecies}) and that storage qubits will rely on individually focused laser beams or ion shuttling for addressed gates (we note this is no different from dual-species operation in this regard).  Since this mode does not rely on an \o\ type transition moment, the ions with very long metastable manifold lifetimes are suitable.

\subsubsection{\{\g,\m,\g\} mode}\label{sec:gmgMode}
{\bf Setting: Fig.~\ref{fig:OMGModes} (middle).}
Individually-addressable, coherent conversion between \g\ and \m\ qubit encodings can be used to activate target ions for \m\ qubit gates.
These gates can then be performed with global beams - a unique advantage of the \omg\ scheme over the dual species scheme. 
Upon gate completion, the active \m\  qubits would be coherently type cast back to \g\ qubits for storage.  Site-specific readout of a multi-ion crystal in the $\{\g,\m,\g\}$ mode would require coherent transfer of all but the target ions to \m\ qubits and read enabling target ions by conversion to \o\ encoding before fluorescence detection.  Similarly, for sympathetic cooling during computation, all logic ions would be type cast to \m\ qubits before cooling. Global readout and cooling at the beginning or end of an algorithm could be done directly.

{\bf Strengths:}
\{\g,\m,\g\} mode provides a high degree of protection of the storage qubits during gates and leverages the stability of \g\ qubits for storage.  Ions with long wavelength \o\ type transitions are particularly well suited to this mode since their \m\ $\leftrightarrow$ \g\ interconversion would be less susceptible to imperfections caused by ion motion than those with short wavelength transitions.  

{\bf Weaknesses:}
\{\g,\m,\g\} mode relies on high-fidelity, coherent transfer on \o\ type transition moments.
This is likely most demanding for site-specific readout or sympathetic cooling during an algorithm, where the type casting of all non-target \g\  qubits to \m\ is necessary to make them insensitive to the dissipative light.

\subsubsection{\{\m,\g,\m\} mode}\label{sec:mgmMode}

{\bf Setting: Fig.~\ref{fig:OMGModes} (bottom).}
Instead of performing gates with \m\ qubits, gates can be performed using \g\ qubits with storage in \m\ encoding in \{\m,\g,\m\} mode.  
For this mode, individually addressed transitions between \g\ $\leftrightarrow$ \m\ would be required to type cast specific ions for gates, cooling, and read enabling, with the latter two possible via incoherent methods.

{\bf Strengths:}
Much like $\{\g,\m,\g\}$ mode, this encoding protects the storage qubits from the laser light used to drive gates, which allows that light to be applied globally. However, unlike $\{\g,\m,\g\}$ mode, only the qubits involved in a process (be it gates, cooling, or readout) need to be interconverted.

{\bf Weaknesses:}
High-fidelity transfer between \g\ and \m\ encodings is required 
Since the $\{\m,\g,\m\}$ modality keeps ions in the metastable states by default, this scheme is limited by the metastable state lifetime and is likely only suitable for species with the longest-lived metastable states.

\subsubsection{Other modes}
Since the utilization of all three qubits encodings is primarily valuable because of its flexibility, it is also possible to envision operational modes that are different from those described above.  For instance, using coherent conversion, the dissipative step in state preparation can be swapped between \m\ and \g\ in any of these modes.  Likewise, the introduction of additional metastable qubit encodings could be used to develop even more flexible schemes that may have advantages along similar lines.  We have chosen to focus on the three above in the interest of clarity, as they illustrate the basic capabilities that flexible \omg\ encodings provide.

\section{Construction and performance of quantum \o \m \g\ primitives}

The operation of an ion processor that takes advantage of the \omg\ architecture will require capabilities that are not yet commonly employed in trapped ion systems. 
In particular, many primitive quantum operations for \m\ qubits are unexplored (though recent work~\cite{yang2021realizing} has begun to address this), and below we outline
four of the important building blocks needed for coherent conversion, state preparation, gates, and to allow for photon-mediated remote entanglement generation.

\subsection{Coherent conversion between \g\ and \m\ encodings}\label{sec:CoherentConversion}

Transformations between \g\ and \m\  qubits can be driven on \o\ type transition moments with two tones of a narrow laser.  By matching the Rabi frequencies for the two transfers ( $\ket{0_{\g}} \leftrightarrow \ket{0_{\m}}$ and $\ket{1_{\g}} \leftrightarrow \ket{1_{\m}}$ ) and setting the frequency difference between the two tones with a stable radio-frequency (rf) source, the laser phase appears globally and the laser does not need to maintain phase coherence between transfers.

Another possibility for coherent $g \! \leftrightarrow \! m$ transfer is via a two-color stimulated Raman transition.  For instance, for $\mathrm{Yb}^+$, an E2+E1 stimulated Raman transition can be driven by combined $411 \mbox{ nm}$ and $3.4\mbox{ }\upmu\text{m}$ light \cite{yang2021realizing}.  While this introduces complexity in the sense that the frequency difference between lasers of dissimilar colors must be stabilized, it has the potential to be a much faster transition than the direct E3 transfer since it does not rely on an E3 transition moment.

\subsection{State preparation and measurement of \m\  qubits}\label{sec:StatePrep}

While state preparation of \g\  qubits for the $\{\g,\m,\g\}$ mode can be accomplished with well established techniques, both the $\{\m,\m,\m\}$ and $\{\m,\g,\m\}$ modes require state preparation of the two \m\  qubit states.
This could be effected by coherent transfer of prepared \g\  qubits to \m\, however, metastable states also offer a simple and robust complementary method.  Unlike \g\ qubits, heralded, probabilistic preparation of \m\  qubits is generally straightforward. Namely, all of the population in undesired states of an initial distribution in the metastable manifold can be state-selectively transferred to the ground state, either by laser-driven transitions on the \o\ type moment or state-selective optical pumping through E1 channels.  Subsequent laser induced fluorescence will herald the preparation of a pure state \m\  qubit when the ion is dark.

Readout is accomplished by moving one of the \m\  qubit states to the \g\ subspace for fluorescence detection of the resulting \o\ manifolds.  This can be done either coherently on the \o\ type transition moment or incoherently through excited E1 channels.  Since this maps the state readout onto discriminating between the \g\ and \m\ manifolds, high fidelity readout should be expected.

\subsection{Coherent gate operations for \m\  qubits}\label{sec:Gates}

Coherent quantum operations on \m\  qubits include single and two-qubit gates.  We describe constructions for these primitives below, including different considerations for two-qubit gates appropriate for ${}^2\mathrm{D}_{5/2}$ and ${}^2\mathrm{F}_{7/2}^o$ state encodings.

\subsubsection{Single \m\ qubit gates}

Single-qubit gates for \m\ qubits can be driven by radiation resonant with the \m\  qubit splitting. Such directly resonant radiation obviates the need for lasers for single-qubit gates, but has the drawback that individual addressing is challenging.   An alternative is to employ laser-driven stimulated Raman transitions, which are discussed below. As shown in Table \ref{table:IonSpecies}, there is a very wide range of \m\ qubit splittings (MHz to GHz) to choose from, so whether one employs directly resonant or Raman transitions, there is an opportunity to take advantage of the potential benefits of matching a particular frequency of qubit to a particular gate approach

\subsubsection{Two \m\ qubit gates in \texorpdfstring{$^2\mathrm{D}_{5/2}$}{doublet D state} ions}

For most of the species in Table \ref{table:IonSpecies}, the \m\  qubit described is encoded in the $^2\mathrm{D}_{5/2}$ manifold. Gates can be performed using rf magnetic fields and gradients or stimulated Raman transitions.  Aside from the lower qubit frequencies in \m\ qubits, magnetic-field-driven gates would proceed much as in \g\  qubits \cite{Ospelkaus08,Mintert2001Ion}; the Raman-beam-driven gates, however, involve different transitions and selection rules than those relevant for \g\  qubits. To achieve a given error in \m\ type Raman gates, somewhat larger detunings and powers are needed compared to requirements calculated in past works~\cite{OzeriRaman2007} for \g\ qubits. This is due to different transition strengths and branching ratios to the qubit manifold; the lack of destructively interfering scattering contributions from ${}^2\mathrm{P}^o_{1/2}$ (since transitions to ${}^2\mathrm{P}^o_{1/2}$ are E1-forbidden from ${}^2\mathrm{D}_{5/2}$); and the smaller Lamb-Dicke parameters (at a given detuning) in \m\ qubits, because of their lower transition frequencies~\cite{DanielRamanNote}.

\subsubsection{Two \m\ qubit gates in \texorpdfstring{$^2\mathrm{F}_{7/2}^o$}{doublet F state} ions}

Gates can be performed using rf magnetic fields and gradients or stimulated Raman transitions. Stimulated Raman transitions for driving gates on the \m\  hyperfine qubit in $\mathrm{Yb}^+$ can be driven by off-resonant E1 coupling to excited states of even parity.  In particular, the $4f^{13}({}^2\mathrm{F}^o_{7/2})6s6p({}^3\mathrm{P}^o_1)\hspace{5pt}(7/2,1)_J$ states are predicted to have a few percent admixture of $6s6p({}^1\mathrm{P}_1)$ character\cite{Biemont1998Lifetime}, and the calculated spontaneous emission probability during an \m\  qubit $\pi$ pulse has a local minimum near $357.2 \mbox{ nm}$.

Alternatively, a $ZZ$ gate scheme \cite{Baldwin2020High} could also be driven by coupling a single qubit state to an excited state.  In particular, the $859.6\mbox{ nm}$ transition to $4f^{13}({}^2\mathrm{F}^o_{7/2})5d6s({}^3\mathrm{D})\hspace{5pt}{}^3[7/2]_{9/2}^o$ falls in a technologically convenient part of the optical spectrum.  However, since the transition moments and branching ratios associated with the excited states have not been measured, some experimental investigation will be needed before appropriate transitions can be confidently identified and compared.

\subsection{Photon-mediated entanglement of \g\  qubits with \omg\ encodings}

Remote entanglement of trapped ions is sometimes performed by heralding an entanglement event after repeated excitation on a strong transition and subsequent interference of the emitted photons\cite{Moehring2007}. This process will lead to decoherence, via ion-ion photon scattering, of any neighboring qubits stored in states that are decayed to on this transition. Utilizing the \omg\ architecture, \m\ qubits can be protected from both laser absorption and ion-ion photon scattering during the \g\ qubit entangling process; entanglement generation trials and quantum logic may even be performed simultaneously on disjoint subsets of qubits when using, e.g., the $\{\m,\m,\m\}$ mode described above. 

\section{Discussion}

The \omg\ architecture implements the idea of utilizing optical, metastable-state, and ground-state qubit encodings within crystals of a \emph{single} ion species, to better optimize state preparation, gates, and readout operations needed for high-performance trapped-ion quantum computation.  
This approach appears to alleviate many of the fundamental challenges encountered when employing multiple ion species and simultaneously reduces the complexity of the requisite optical technology.
Some potential challenges arising in the \omg\ paradigm are the possibility of ac Zeeman shifts of metastable qubits due to the trap rf field (both can be in the tens of megahertz range) and the increased need for tightly focused beams, or ion shuttling, to enable individual addressing.  

The new primitives and modes of operation described above for metastable-state qubits may be extended to quantum information processing applications not considered here (e.g.\ to enable a subset of monitor qubits used for calibration or sensing of environmental noise during other operations).  Beyond the specific ideas detailed here, further utility could be gained from employing more than two hyperfine levels for additional quantum state encodings.  Improved gates on \omg\ qubits may also be realized using pulsed (instead of CW) lasers, in some cases at much more technologically favorable wavelengths.
These options indicate a rich \omg\ future for trapped-ion quantum-computer architectures.

\section*{Acknowledgements}
This work was supported in part by the US Army Research Office under award W911NF-20-1-0037.  I.L.C.\ acknowledges support by the NSF Center for Ultracold Atoms.  D.T.C.A., I.D.M., and D.J.W. wish to acknowledge support from NSF through the Q-SEnSE Quantum Leap Challenge Institute, Award \# 2016244.  W.C.C. and E.R.H. acknowledge support from the CIQC Quantum Leap Challenge Institute through NSF award OMA-2016245 and W.C.C. under award PHY-1912555.

\section*{Author Declarations}
The authors have no conflicts to disclose.

\section*{Data Availability}
Data sharing not applicable – no new data generated.


\bibliography{OMGBib}

\end{document}